\documentstyle[aps]{revtex}

\begin{document}
\draft
\date{Submitted to Physical Review B, 31 May 1996}

\title {Fine structure of excitons in Cu$_2$O}
\author{G. M. Kavoulakis, Yia-Chung Chang, and Gordon Baym}
\address {Department of Physics, University of Illinois at
 Urbana-Champaign, 1110 West Green Street, Urbana, Illinois 61801}
\maketitle

\begin{abstract} 
\baselineskip=14pt

    Three experimental observations on 1s-excitons in Cu$_2$O are not
consistent with the picture of the exciton as a simple hydrogenic bound state:
the energies of the 1s-excitons deviate from the Rydberg formula, the total
exciton mass exceeds the sum of the electron and hole effective masses, and
the triplet-state excitons lie above the singlet.  Incorporating the band
structure of the material, we calculate the corrections to this simple picture
arising from the fact that the exciton Bohr radius is comparable to the
lattice constant.  By means of a self-consistent variational calculation of
the total exciton mass as well as the ground-state energy of the singlet and
the triplet-state excitons, we find excellent agreement with experiment.

\end{abstract}

\baselineskip=14pt
\section{Introduction}

     The absorption spectrum of light in Cu$_2$O \cite{OPA} shows clear
evidence for the existence of excitons. In the simplest picture the
exciton is described as a hydrogenic atom formed from electrons and holes
of given effective masses interacting through a Coulomb interaction
modified by a dielectric constant \cite{Knox}. Three observations indicate
that this simple picture needs to be refined: 1) The exciton Rydberg,
$\mu e^4/2 \hbar^2 \epsilon_0^2$ (where $\mu$ is the reduced electron-hole 
mass and $\epsilon_0$ is the static dielectric constant of the material)
is 98 meV; experimentally it is measured to be 97 meV
for the $n=2,3,...$ states, however, for the 1s-state it has the anomalously
high value of 153 meV \cite{Knox}. 2) The mass of the lowest (yellow series)
$^3$S$_1$
(ortho) exciton is experimentally $M_o=(3.0 \pm 0.2)m$ \cite{Yu,Snoke},
where $m$ is the free-electron mass;
on the other hand the sum of the electron effective mass 
$m_e=(0.99 \pm 0.03)m$ and hole mass $m_h=(0.69 \pm 0.04) m$ \cite{Hodby}
is only $m_e + m_h = (1.68 \pm 0.07)m$. 3) In a simple hydrogenic model
the ortho and para ($^1$S$_0$) excitons would be degenerate.
However the lowest ortho excitons lie 12 meV higher than the lowest
para excitons \cite{exch}.

    Our purpose in this paper is to identify the
salient physics responsible for these observations. These are several
effects. First, one must take into account the non-parabolicity of the
bands \cite{Yu}. Furthermore, the electron-hole interaction is
more properly the bare Coulomb interaction modified by the
{\it momentum} and {\it frequency} dependent dielectric function
\cite{Haken,Makarov}. In addition, 
the spin-dependent exchange interaction between the electron and the hole 
lifts the degeneracy between the triplet and 
the singlet-state excitons \cite{exch}.
All these effects become important because the Bohr radius $a_B$ is not large
compared with the lattice constant $a_{\ell}$ of the material \cite{Bassani}.
For Cu$_2$O, $a_{\ell} = 4.26$ \AA, while for the 1s-state 
yellow excitons the Bohr radius $a_B$ is expected to be on the order of
$e^2/(2 \epsilon_0 E_b) \approx 7$ \AA, where $E_b$ is the {\it observed}
binding energy, $\approx$ 153 meV. Because the Bohr radius
increases quadratically with the principal quantum number $n$, these
effects are much more important for the $n = 1$ state.

      The correction due to the non-parabolicity of the bands is
expected to be significant, since the extent of the exciton
wavefunction in momentum space is of order $1/a_B$,
while the width of the Brillouin zone is of order $1/a_{\ell}$.
This correction makes the exciton heavier, since
away from the zone center and closer to the edge of the Brillouin
zone, the dispersion relation of the bands flattens and the 
electron and hole bare-band masses effectively increase.

       The coupling of the electron and the hole to the
LO-phonons produces a frequency dependence of the dielectric
function $\epsilon(k,\omega)$ on the scale of the phonon frequencies.
In the limit that the frequency of
the relative electron-hole motion is much larger than that of the LO-phonon,
the electron-hole interaction is screened by the high-frequency dielectric
constant, $\epsilon_{\infty}$, since the heavy ions cannot follow the
motion of the electron and hole and therefore they do not contribute
to the screening. In the opposite limit the low-frequency
dielectric constant $\epsilon_0$ screens the electron-hole interaction.
When the Bohr radius is comparable to the lattice constant $a_{\ell}$,
the momentum dependence of the dielectric function, on scales of
$\hbar/a_{\ell}$, becomes important. The more localized 1s-exciton
states are screened by $\epsilon$ at higher momenta, making the effective
Coulomb interaction stronger than for the larger excited exciton states.
Finally the exchange interaction is short-ranged and is negligible
for excitons with $a_B \gg a_{\ell}$. All these corrections, 
known as the ``central-cell corrections,''
act to produce the fine structure of excitons.

     Cuprus oxide has in total ten valence and four conduction bands. It
has a direct gap, since the minimum of the lowest
conduction band ($\Gamma_6^+$) is at the same point 
in momentum space as the maximum  of the 
highest valence band, ($\Gamma_7^+$); the gap energy is $\approx$ 2.17 eV. See
Fig. 1. The yellow-series excitons are formed between electrons and holes in
these two bands. Since the conduction
and valence bands have the same (positive) parity \cite{bstr} and the 
dipole moment between them vanishes, the radiative lifetimes of the excitons 
are relatively long. The $n=1$ line in the one-photon absorption spectrum
of light is weak due to the equal parity of the conduction and valence bands; 
the $n \neq 1$ lines
correspond to excitons with relative angular momentum $l=1$ and for this
reason the absorption process is dipole-allowed.
The electrons in the $\Gamma_7^+$ band are not in pure spin states, but rather 
in total angular-momentum states; the direct recombination
process of the angular-momentum singlet-state para exciton
is in fact highly forbidden, and the corresponding line is absent
from the radiative recombination spectrum of
Cu$_2$O. The lower $\Gamma_8^+$ valence band, which lies 
$\approx 130$ meV below the $\Gamma_7^+$ band due to the spin-orbit 
interaction (Fig. 1), forms, with the $\Gamma_6^+$ band, the green-exciton
series. Here we neglect for simplicity any possible
mixing between the yellow and the green-exciton series. This mixing is expected
to be on the order of 10\% \cite{Trebin}, and its only result is to modify 
slightly the exciton binding energies. 

    Recently high-density excitons in Cu$_2$O have been observed to 
obey Bose-Einstein statistics \cite{HMB,SWM87,augo} and indeed  
Bose-Einstein condensation \cite{Griffin} has been observed \cite{SWM90,LW}.
These observations are directly related with the band structure of Cu$_2$O, 
as we have shown in Refs. \cite{kbw,KB}.

   In this paper we start with the effective mass approximation, which
we describe in Sec. II. In Sec. III we discuss the central-cell corrections.  
In Sec. IV we study the exchange interaction \cite{exch}, and review
the band structure which underlies the  
properties of excitons in Cu$_2$O \cite{Nik}.
We summarize our results in Sec. V.

\section{Effective mass approximation for excitons} 

      An exciton in the effective mass approximation is a hydrogen-like
bound state of an electron and a hole, with center-of-mass 
in a plane-wave state. The exciton energies lie in discrete levels 
below the energy gap, 
determined by the binding energy plus the energy carried by the
center-of-mass. In the effective mass approximation the Hamiltonian $H$ 
of an electron and a hole which interact through their Coulomb attraction, 
modified by a dielectric constant $\epsilon_0$, is
\begin{eqnarray}
     H = \frac {p_e^2} {2 m_e} 
   + \frac {p_h^2} {2 m_h} 
   - \frac {e^2} {\epsilon_0 |{\bf r_e - r_h}|},
\label{21}
\end{eqnarray}
where ${\bf p_i}$ is the momentum operator of the electron and the hole
and the $m_i$ are the effective electron and hole masses. 
The Hamiltonian can be written in terms of the momentum and the coordinate
operators of the relative motion of the electron and the hole, ${\bf p}$ and 
${\bf r}$ respectively, and the momentum operator ${\bf P}$ of the 
center-of-mass as
\begin{eqnarray}
    H = \frac {P^2} {2 M} 
   + \frac {p^2} {2 \mu} - \frac {e^2} {\epsilon_0 r},
\label{22}
\end{eqnarray}
where $M=m_e+m_h$ is the total exciton mass and $\mu=m_e m_h/(m_e + m_h)$
is the reduced mass. The eigenfunctions of the above Hamiltonian are of 
the form
\begin{eqnarray}
     \Psi_{{\bf K},nlm}({\bf r},{\bf R})=
    \frac 1 {\sqrt \Omega} e^{i {\bf K} \cdot {\bf R}}
   \Phi_{nl}(r) Y_l^m(\theta,\phi),
\label{23}
\end{eqnarray}
where $\Omega$ is volume of the crystal,
the $Y_l^m$ are spherical harmonics and
$\Phi_{nl}$ are the radial hydrogenic eigenfunctions.
In the state (\ref{23})
the center-of-mass of the exciton carries momentum $\hbar \bf{K}$.
The corresponding eigenenergies are
\begin{eqnarray}
   E_{{\bf K},n} = E_g + \varepsilon^{(0)}_n + \frac {\hbar^2 K^2} {2 M},
\label{24}
\end{eqnarray}
where $\varepsilon^{(0)}_n = -\mu e^4/2 \hbar^2 \epsilon_0^2 n^2$
and $E_g$ is the band-gap energy.

    If we assume for simplicity that the effective electron and hole masses are 
equal, the exciton wavefunction can be expressed as a linear superposition of
electron and hole Bloch states as
\begin{eqnarray}
    \Psi_{{\bf K},nlm}({\bf r_e},{\bf r_h})= \sum_{{\bf q}} \phi_{{\bf q}} \,
   \Phi_{c,{\bf q} + {\bf K}/2}({\bf r_e}) \Phi_{v,-{\bf q} + {\bf K}/2}
  ({\bf r_h}),
\label{25}
\end{eqnarray}
where the Bloch states are of the usual form
\begin{eqnarray}
  \Phi_{j,{\bf k}}({\bf r}) = u_{j,{\bf k}}({\bf r}) e^{i {\bf k} \cdot 
 {\bf r}},
\label{26}
\end{eqnarray}
with $u_{j,{\bf k}}({\bf r})$ periodic.
To a good approximation we can identify $\phi_{{\bf q}}$ as the Fourier
transform of the relative electron-hole wavefunction times $\sqrt {\Omega}$.

\section{Central--cell corrections}

    As we mentioned earlier, the 
binding energy of excitons is modified by the non-parabolicity of the bands
\cite{Yu}, the coupling of the electron and the hole with the
longitudinal-optical (LO) phonons \cite{Haken},
the dependence of the dielectric function on the distance between
the electron and the hole \cite{Makarov} and the exchange interaction
\cite{exch}. In this section we study the first three mechanisms.

     Let us start with the correction due to the non-parabolicity of the bands. 
In the tight-binding approximation, for example, the electron or hole  
dispersion relation is a cosine function of $k a_{\ell}$.
Expanding this function around zero, the zero-order term gives a
constant. The first correction, $\sim (k a_{\ell})^2$, describes a free
particle with an effective band mass $m_{i,b}$. The next non-vanishing
term is of the form 
\begin{eqnarray}
     \Delta V_i =  \frac {p_i^4 a_{\ell}^2} {24 \hbar^2 m_i}.
\label{2223i}
\end{eqnarray}
More generally, the perturbation $\Delta V$ due to the non-parabolicity
of the Hamiltonian is of the form,
\begin{eqnarray}
     \Delta V =  \frac {p_e^4 a_{\ell}^2 C_e} {24 \hbar^2 m_e} 
       + \frac {p_h^4 a_{\ell}^2 C_h} {24 \hbar^2 m_h},
\label{223i}
\end{eqnarray}
where the constants $C_i$ are on the order of unity.
We proceed by writing Eq. (\ref{223i}) in terms of the center-of-mass
coordinates, keeping terms up to order $p^2 P^2$,
\begin{eqnarray}
 \Delta V = \Delta V_{\alpha} + \Delta V_{\beta} \equiv
      \frac {p^4 a_{\ell}^2} {24 \hbar^2 \mu'} -
     \frac {p^2 P^2 a_{\ell}^2} {4 \hbar^2 M'},
\label{231}
\end{eqnarray}
where $1/\mu' \equiv (C_e/m_e)+(C_h/m_h)$ and $M' \equiv (m_e/C_e) + (m_h/C_h)$.
The $\Delta V_{\alpha}$ in Eq. (\ref{231}) refers to the
relative electron-hole motion; the second couples the relative motion 
of the electron and hole with the motion of their
center-of-mass and modifies the
total exciton mass. We treat $\Delta V_{\alpha}$
as a perturbation in calculating its contribution to the binding energy.
For the central-cell corrections 
we use a trial hydrogen-like wavefunction of the form $\Psi(r) =
1/(\pi a_B^3)^{1/2} e^{-r/a_B}$ but we truncate it outside the first
Brillouin zone, assuming that $\Psi_{\bf q}$, the Fourier transform of 
$\Psi(r)$, vanishes for $|{\bf q}| > \pi / a_{\ell}$.
The reason for truncating $\Psi_{\bf q}$ is that  
for values of the Bohr radius smaller or comparable to $a_{\ell}$,
the exciton wavefunction is spread in momentum space and therefore
without the truncation the central-cell corrections are overestimated.
The expectation value of $\Delta V_{\alpha}$ in $\Psi_{\bf q}$ is,
\begin{eqnarray}
  \langle \Delta V_{\alpha} \rangle =
   - \frac {\hbar^2 a_{\ell}^2} {24 \mu'}
   \left( \sum_{|{\bf q}| < \pi / a_{\ell}} |\Psi_{\bf q}|^2 q^4 \right)
      \left( \sum_{|{\bf q}| < \pi / a_{\ell}} |\Psi_{\bf q}|^2 \right)^{-1}
	= - \frac {\hbar^2 a_{\ell}^2} {24 \mu a_B^4}
         \, \frac {m_h C_e + m_e C_h} {m_e + m_h} \,
	       \frac {I_6(\pi a_B/a_{\ell})} {I_2(\pi a_B/a_{\ell})},
\label{2321}
\end{eqnarray}
where 
\begin{eqnarray}
   I_n(x) = \int_0^x \frac {y^n dy} {(1+y^2)^4}.
\label{integral}
\end{eqnarray}
In the limit $a_B \gg a_{\ell}$, Eq. (\ref{2321}) gives the result
\begin{eqnarray}
  \langle \Delta V_{\alpha} \rangle =
    - \frac {5 \,\hbar^2 a_{\ell}^2} {24 \mu a_B^4}
     \, \frac {m_h C_e + m_e C_h} {m_e + m_h}.
\label{232}
\end{eqnarray}

     The second term of Eq. (\ref{231}), $\Delta V_{\beta}$,
is the first correction to the
total exciton mass due to the non-parabolicity of the bands. This term
modifies the free dispersion relation for the center-of-mass motion to
\begin{eqnarray}
   E_{{\bf K}} = \frac {\hbar^2 K^2} {2 M} +
\frac {\langle p^2 \rangle a_{\ell}^2} {4 M'} K^2
  = \frac {\hbar^2 K^2} {2 M}
   \left( 1 - \frac {M a_{\ell}^2} {2 M' a_B^2} 
   \frac {I_4(\pi a_B/a_{\ell})} {I_2(\pi a_B/a_{\ell})} \right),
\label{234}
\end{eqnarray}
making the total exciton mass larger than $M$. The above expression 
is in fact the total exciton mass, since the contribution of the exchange 
interaction to the mass is negligible, as shown in Sec. IV.
Again, in the limit $a_B \gg a_{\ell}$, Eq. (\ref{234}) gives the result
\begin{eqnarray}
   E_{{\bf K}} = 
      \frac {\hbar^2 K^2} {2 M}
	\left( 1 - \frac {M a_{\ell}^2} {2 M' a_B^2} \right).
\label{2341}
\end{eqnarray}

    We estimate now the order of magnitude of the constants $C_i$,
using ${\bf k} \cdot {\bf p}$ perturbation theory.
Among the ten valence and four conduction bands of Cu$_2$O only one $\Gamma_8^-$
conduction band, which lies $\approx$ 449 meV above the $\Gamma_6^+$ band,
and one very deep $\Gamma_8^-$ valence band, which lies $\approx$ 5.6 eV
below the $\Gamma_7^+$ valence band, have
negative parity. The mixing of the $\Gamma_6^+$ and
$\Gamma_7^+$ bands with these two negative-parity bands modifies the
bare masses of the electrons and the holes (in addition to the coupling
of the electrons and the holes to the optical modes)
and also makes the bands non-parabolic. To find the constants $C_i$
in lowest order requires diagonalizing
a 4$\times$4 matrix; here we give just an order of magnitude estimate.
The correction to the masses is of order $|{\bf p}_{i,j}|^2/
m \Delta_{i,j}$, where ${\bf p}_{i,j}$ is the dipole matrix element between
the opposite-parity bands $i$ and $j$ and $\Delta_{i,j}$ is the energy 
separation between them. The matrix elements ${\bf p}_{i,j}$ can be extracted 
from experiment; if we assume, for example, that they are all of equal 
magnitude, then from Ref. \cite{Nik}, $|{\bf p}_{i,j}|/\hbar \approx 
0.13$ \AA$^{-1}$, which implies that $|{\bf p}_{i,j}|^2/ m \Delta_{i,j} 
\approx 0.3$, for $\Delta_{i,j} \approx$ 0.5 eV. Dimensionally  
we find 
\begin{eqnarray}
      \frac {\hbar^2 a_{\ell}^2 C k^4} {24 m} \sim
     \frac {\hbar^2 k^2} {2m}
    \frac {|{\bf p}_{i,j}|^4} {m^2 \Delta_{i,j}^2}
   \frac {\hbar^2 k^2} {m \Delta_{i,j}},
\label{exp}
\end{eqnarray}
which gives $C \sim 1$.

     The Hamiltonian (\ref{21}) describes the effective interaction
between an electron and a hole at momentum transfer $k$ as
\begin{eqnarray}
    V(k) = \frac {4 \pi e^2} {k^2 \epsilon_0}.
\label{inter1}
\end{eqnarray}
More generally, the interaction is given by 
\begin{eqnarray}
  V(k,\omega) = \frac {4 \pi e^2} {k^2 \epsilon(k,\omega)},
\label{inter2}
\end{eqnarray}
where $\epsilon(k,\omega)$ is the momentum- and frequency-dependent 
dielectric function, and $\hbar k$ and $\hbar \omega$
are the momentum and energy transferred in the interaction.
The coupling of the electron and the hole to the
LO-phonons introduces the important frequency dependence $\omega$ in the 
dielectric function \cite{Haken}.
In Cu$_2$O only the $\Gamma_{15}^{-}$ LO-phonon modes with zone-center
energies of 18.7 and 87 meV contribute to the Fr\"ohlich interaction.
For $\hbar \omega \ll 18.7$ meV, the dielectric function has the
low-frequency value, $\epsilon_0=7.5 \pm 0.2$. As the $\hbar \omega$
crosses 18.7 meV, $\epsilon$ decreases to $\epsilon_m = 7.11$, while
as $\hbar \omega$ increases past 87 meV, $\epsilon$ drops to 
$\epsilon_{\infty}=6.46$. Comparing the LO-phonon
frequencies with the frequency of the relative electron-hole motion, we see 
that the dielectric constant of the 1s-state is
$\, {\raisebox{-.5ex}{$\stackrel{<}{\sim}$}} \, \epsilon_{\infty}$,
while for the excited $n=2,3,4,...$ states the dielectric
constant is closer to $\epsilon_{0}$. 
Using the values for the effective electron and
hole masses given in the introduction and $\epsilon_{0}=7.5$  
we find 98 meV for the Rydberg of the excited states, 
which is very close to the experimentally determined value of 97 meV.
Assuming that $\epsilon_{\infty}$ screens the electron-hole interaction,
the expectation value $\langle PE \rangle$, of the Coulomb interaction in 
the 1s-state is
\begin{eqnarray}
     \langle PE \rangle = - \frac {e^2} {\epsilon_{\infty} \, a_B}.
\label{PE}
\end{eqnarray}
Since the actual dielectric constant is
$\, {\raisebox{-.5ex}{$\stackrel{<}{\sim}$}} \, \epsilon_{\infty}$,
the choice of $\epsilon_{\infty}$ overestimates the correction
to the potential energy due to the coupling of the electrons and 
holes to the optical modes of the crystal.
Although there are more
sophisticated methods of treating this problem \cite{Haken}, they cannot 
be applied in our problem because more than one LO-phonon branches
contributes to the Fr\"ohlich interaction.

   We turn now to the effects of the momentum dependence of the dielectric 
function. Equation (\ref{inter2}) gives the interaction between the electron
and the hole, where as discussed above the high-frequency
dielectric function must be used. 
In the limit $a_B \gg a_{\ell}$, $\Psi_{{\bf k}}$ is localized 
around zero, so we can ignore the $k$-dependence of the dielectric function,
$\epsilon(k) \approx \epsilon(k=0) \approx \epsilon_{\infty}$.
On the other hand, if the Bohr radius is comparable to
the lattice constant, we need to consider corrections to $\epsilon(k)$.
For small values of $k$
\begin{eqnarray}
      \epsilon(k) \approx \epsilon_{\infty} - (k a_{\ell})^2 \,d,
\label{225}
\end{eqnarray}
where, as we calculate below, $d \approx 0.18$ for Cu$_2$O.
Expanding, 
\begin{eqnarray}
    \frac {4 \pi e^2} {k^2 \epsilon(k)} \approx 
     \frac {4 \pi e^2} {k^2 \epsilon_{\infty}}
      + \frac {4 \pi e^2} {\epsilon_{\infty}^2} a_{\ell}^2 d,
\label{expan}
\end{eqnarray}
we see that the first-momentum correction to $\epsilon(k)$ produces an
effective contact interaction 
\begin{eqnarray}
     \Delta V_d ({\bf r}) = - \frac {4 \pi e^2} {\epsilon_{\infty}^2}
   d a_{\ell}^2
   \, \delta({\bf r}),
\label{contact}
\end{eqnarray}
which lowers the exciton energy by 
\begin{eqnarray}
   \langle \Delta V_d \rangle =
      - \frac {4 \pi e^2} {\epsilon_{\infty}^2} d
         a_{\ell}^2 |\Psi_{\text{exact}}(0)|^2 ,
\label{contcor}
\end{eqnarray}
where $\Psi_{\text{exact}}(r)$ is the exact (unknown) wavefunction of the 
1s-state. If we evaluate $\Delta V_d$ with the trial wavefunction 
$\Psi(r)$ truncated outside the first Brilloin zone we find
\begin{eqnarray}
      \langle \Delta V_d \rangle = - \frac {4 \pi e^2 d a_{\ell}^2}
     {\Omega \epsilon_{\infty}^2}
    \left| \sum_{|{\bf q}| < \pi / a_{\ell}} \Psi_{\bf q} \right|^2 \left(
\sum_{|{\bf q}| < \pi / a_{\ell}} |\Psi_{\bf q}|^2 \right)^{-1}
  =- \frac {2 d e^2 a_{\ell}^2} {\pi \epsilon_{\infty}^2 a_B^3}
       \frac {[ I_2^{'}(\pi a_B/a_{\ell})]^2} {I_2(\pi a_B/a_{\ell})},
\label{dielectric}
\end{eqnarray}
where $\Psi_{\bf q}$ is again the Fourier transform of $\Psi(r)$.
Also
\begin{eqnarray}
       I_n^{'}(x) = \int_0^x \frac {y^n dy} {(1+y^2)^2}.
\label{integrall}
\end{eqnarray}
In the limit $a_B \gg a_{\ell}$, Eq. (\ref{dielectric}) gives the
result
\begin{eqnarray}
	\langle \Delta V_d \rangle =
     - \frac {4 \pi e^2} {\epsilon_{\infty}^2} d
	a_{\ell}^2 |\Phi_{1s}(0)|^2 = - \frac {4 d e^2 a_{\ell}^2}
   {\epsilon_{\infty}^2 a_B^3},
\label{228}
\end{eqnarray}

    To estimate the constant $d$ we follow Ref. \cite{Hermanson}.
The Lindhard result gives for the dielectric function
\begin{eqnarray}
        \epsilon(k,\omega) = 1 + \frac {4 \pi e^2} {\Omega_c k^2}
   \sum_{{\bf q},l,l'} 
   \frac {| \langle u_{l',{\bf k} + {\bf q}} 
   \mid u_{l, {\bf q}} \rangle |^2} 
  {\varepsilon_{l',{\bf k} + {\bf q}} - \varepsilon_{l,{\bf q}} - \hbar \omega},
\label{Lind}
\end{eqnarray}
where $\Omega_c$ is the volume of the unit cell and
$\varepsilon_{l,{\bf q}}$ is the energy of band $l$ at point 
${\bf q}$ in momentum space. Since we are interested in energies much smaller
than the gap energy, we assume that $\omega = 0$.
The dominant contribution to the above sum,
given the band structure of Cu$_2$O, involves virtual transitions 
between the $\Gamma_8^{-}$ conduction band $c'$ (which lies
$\approx$ 449 meV above the lowest $\Gamma_6^{+}$ conduction band) and the
highest $\Gamma_7^{+}$ valence band $v$. The overlap integral in 
Eq. (\ref{Lind}) satisfies the sum rule \cite{Pines}
\begin{eqnarray}
       \sum_{l'} | \langle u_{l',{\bf k} + {\bf q}} 
   \mid u_{l, {\bf q}} \rangle |^2
  ( \varepsilon_{l',{\bf k} + {\bf q}} - \varepsilon_{l,{\bf q}} ) =
    \frac {\hbar^2 k^2} {2 m},
\label{sumrule}
\end{eqnarray}
which as in Ref. \cite{Hermanson} allows us to write 
the approximate result for $\epsilon(k)$,
with $\bf k$ restricted in the first zone,
\begin{eqnarray}
    \epsilon(k) \approx 1 + \frac {4 \pi e^2 \hbar^2} {m \Omega_c}
   \sum_{{\bf q}} (\varepsilon_{c',{\bf k} + {\bf q}} - 
  \varepsilon_{v,{\bf q}} )^{-2}.
\label{approx}
\end{eqnarray}
If we define
\begin{eqnarray}
    \Sigma(k) \equiv \sum_{{\bf q}} (\varepsilon_{c',{\bf k} + {\bf q}} -
      \varepsilon_{v,{\bf q}} )^{-2},
\label{sigma}
\end{eqnarray}
the dielectric function can be written as 
\begin{eqnarray}
     \epsilon(k) = 1 + (\epsilon_{\infty} - 1) \Sigma(k) / \Sigma(0).
\label{interpol}
\end{eqnarray}
We calculate the above quantity numerically, using the effective mass of the
conduction $\Gamma_8^{-}$ band $m_e^{'} = 0.35m$, the effective mass of the 
valence $\Gamma_7^{+}$ band $m_h = 0.69m$ and the energy gap between them,
$\approx$ 2.62 eV. The integration over $\bf q$ is restricted to $q \le k_D$,
where $k_D^3=6 \pi^3/\Omega_c$.
For $k \rightarrow 0$, Eq. (\ref{interpol}) is of the form of Eq. (\ref{225})
with $d \approx 0.18$.

   Equations (\ref{2321}), (\ref{PE}) and (\ref{dielectric}) give
the total exciton energy as function of the Bohr radius,
\begin{eqnarray}
   E_{\text{tot}}(a_B) = \frac {\hbar^2} {2 \mu a_B^2} 
       -\frac {e^2} {\epsilon_{\infty} a_B}
	- \frac {C \, \hbar^2 a_{\ell}^2} {24 \mu a_B^4}
           \frac {I_6(\pi a_B/a_{\ell})} {I_2(\pi a_B/a_{\ell})}
     - \frac {2 \, d e^2 a_{\ell}^2} {\pi \epsilon_{\infty}^2 a_B^3}
	  \frac {[ I_2^{'}(\pi a_B/a_{\ell})]^2} {I_2(\pi a_B/a_{\ell})}.
\label{total}
\end{eqnarray}
The first term is the kinetic energy of the electron-hole
pair, $\langle KE \rangle$. The second term is the potential energy 
$\langle PE \rangle$, 
the third term is the correction due to the non-parabolicity
of the bands, $\langle \Delta V_a \rangle$, and the last term is the 
correction due to the dependence of the dielectric 
function on the momentum, $\langle \Delta V_d \rangle$. 
In the third term we have made the simplifying assumption
$C_e=C_h=C$. For given $C$ the binding energy has a minimum at 
$a_B=a_{B,0}$, which is the exciton Bohr radius for the specific choice
of $C$. The corresponding binding energy is $E_{\text{tot}}(a_{B,0})$.
The total exciton mass is then given by Eq. (\ref{234}) with $a_B=a_{B,0}$.
Equation (\ref{exp}) gave us an order of magnitude 
estimate $C \sim 1$. The value $C=1.45$ gives the result $a_{B,0} = 5.3$ \AA \,
and the observed total ortho-exciton mass of $3m$. 
This choice for $C$ then yields
$E_{\text{tot}}(a_{B,0}) = 166.9$ meV; see Fig. 2, in good agreement (10\%)
with the experimentally known binding energy of $153$ meV.
The expectation value of each of the terms separately in (\ref{total}) is:
$\langle PE \rangle = -420.4$ meV, $\langle KE \rangle = 335.8$ meV,
$\langle \Delta V_a \rangle = -68.3$ meV and $\langle \Delta V_d
\rangle = -14.0$ meV. We expect to find a larger binding energy here
than the experimental value, since the choice of
$\epsilon_{\infty}$ for the dielectric constant overestimates the 
potential energy. The Bohr radius used in the literature for excitons 
in Cu$_2$O is 7 \AA, the number resulting from the uncorrected  
formula $a_B = e^2/(2 \epsilon_0 E_b)$ with $E_b =153$ meV,
the observed binding energy. The value of $E_b= 98$ meV, which 
is the Rydberg of the excited states, gives $a_B \approx 11.1$ \AA \,
in this way. But as we have seen the formula $a_B = e^2/(2 \epsilon_0 E_b)$ 
neglects the central-cell corrections.

    The total mass of the $n=1$ para excitons must be the same
as that of the ortho excitons, since the correction due to the 
non-parabolicity of the bands is the same for both the singlet and the triplet;
the correction due to the exchange interaction which is nonzero only for the
ortho excitons is negligible in our problem. Furthermore,
the quadratic dependence of the Bohr radius on the quantum number $n$
implies that the total exciton mass
is very close to $m_e+m_h \approx 1.68m$ for the excited states.
Neither the $n=1$ para-exciton mass nor the $n \ge 2$ exciton masses 
has yet been measured.

\section{Band structure of Cu$_2$O -- 
        Exchange interaction and optical properties of excitons in Cu$_2$O}

    The virtual annihilation of an exciton, shown in Fig. 3, is 
responsible for raising the ortho exciton by an energy 
$\Delta E_{\text{ex}}=12$ meV above the para exciton 
at the zone center ($K=0$) of Cu$_2$O \cite{exch}. 
The energy splitting $\Delta E_{\text{ex}}(K)$ is given by
\begin{eqnarray}
      \Delta E_{\text{ex}}(K)=\frac 2 3 \int \Psi_{{\bf K}}({\bf r},{\bf r}) 
     \frac {e^2} {\epsilon_{\infty} |{\bf r} - {\bf r '}|}
   \Psi_{{\bf K}}^{*}({\bf r '} , {\bf r '}) \, d {\bf r} \, d {\bf r'}.
\label{221}
\end{eqnarray}
The factor 2/3 comes from the angular momentum states, as we show in
the next section. If the electrons and the holes in the conduction and valence
bands are in pure-spin states, the above interaction is non-zero for the singlet
excitons only. Since it is positive, it shifts the energy of the singlet
higher than the triplet. In the next section
we explain how the band structure of Cu$_2$O makes the exchange
interaction non-zero for the ortho excitons and zero for the para excitons,
shifting the ortho excitons higher in energy than the para excitons.
Using Eq. (\ref{25}) for the exciton wavefunction we find that
\begin{eqnarray}
    \Delta E_{\text{ex}}(K) &=& \frac 2 3 \,
    \frac 1 {\Omega_c} \sum_{{\bf q},{\bf p},{\bf G} \neq 0} 
      \frac {4 \pi e^2}
    {|{\bf G}|^2 \epsilon_{\infty}}
    \phi_{{\bf q}} 
   \langle u_{v,{\bf q}} | u_{c,{\bf q}} \rangle_{{\bf G}} \,
  \phi_{{\bf p}}^{*} 
    \langle u_{v,{\bf p}} | u_{c,{\bf p}} \rangle_{{\bf G}}^*
\nonumber \\
             &+& \frac 2 3 \, \frac 1 {\Omega_c}
           \sum_{{\bf q},{\bf p}} \frac {4 \pi e^2}
         {|{\bf K}|^2 \epsilon_{\infty}}
       \phi_{{\bf q}}
\langle u_{v,{\bf q - K/2}} | u_{c,{\bf q + K/2}} \rangle \,
        \phi_{{\bf p}}^{*} 
\langle u_{v,{\bf p - K/2}} | u_{c,{\bf p +K/2}} \rangle^*,
\label{222}
\end{eqnarray}
where
\begin{eqnarray}
    \langle u_{v,{\bf q}} | u_{c,{\bf p}} \rangle_{{\bf G}}=
       \int d {\bf x} \, u_{c,{\bf p}}({\bf x})
	    u_{v,{\bf q}}^{*}({\bf x}) e^{i {\bf G} \cdot {\bf x}}.
\label{defn}
\end{eqnarray}
The first sum in Eq. (\ref{222}) is over all the non-zero 
reciprocal lattice vectors ${\bf G}$ 
of the crystal. In this equation we have used the high-frequency dielectric
constant in the Coulomb interaction 
because the energy transfer in the virtual annihilation 
process of an exciton is on the order
of the energy gap $E_g$. As we show below, the overlap integrals
\cite{overl,Land} which appear
in the second term of Eq. (\ref{222}) are proportional to $(K^2)^2$ because
the dipole moment between the conduction and the valence bands vanishes.
The second term of (\ref{222}) therefore goes as $K^2$, since it is
proportional to $1/K^2$ from the Coulomb 
interaction times $(K^2)^2$ from the overlap integrals; it vanishes at the
zone center and renormalizes the ortho-exciton mass.
To calculate $\Delta E_{\text{ex}}$ at $K=0$, we assume that
the first sum of Eq. (\ref{222}) is dominated by the terms 
with smallest ${\bf G}$'s 
(six in number because of the cubic symmetry of the crystal),
which we denote by $G_0$; then
\begin{eqnarray}
       \Delta E_{\text{ex}}(K=0) \, {\raisebox{-.5ex}{$\stackrel{<}{\sim}$}} \,
        \frac {16 \pi e^2} {\Omega_c \epsilon_{\infty}}
      \Omega_c |\Phi_{1s}(0)|^2 \frac {a_{\ell}^2} {4 \pi^2}
      |\langle u_{v,{\bf 0}} | u_{c,{\bf 0}} \rangle_{{\bf G_0}}|^2
     =  \frac {e^2} {a_B \epsilon_{0}} 
    \frac 4 {\pi^2} \left( \frac {\epsilon_{0}} {\epsilon_{\infty}} \right)
   \left( \frac {a_{\ell}} {a_B} \right)^2
   |\langle u_{v,{\bf 0}} | u_{c,{\bf 0}} \rangle_{{\bf G_0}}|^2.
\label{223}
\end{eqnarray}
Since experimentally $\Delta E_{\text{ex}}(K=0) \approx 12$ meV, we have
$|\langle u_{v,{\bf 0}} | u_{c,{\bf 0}} \rangle_{{\bf G_0}}| \approx 0.45$.

    In most semiconductors the dipole matrix element ${\bf p}_{c,v}$ between the
conduction and the valence bands does not vanish. In this case the second term 
in Eq. (\ref{222}) is proportional to $1/K^2$ from the Coulomb interaction
times $K^2$ from the overlap integrals and therefore from ${\bf k} \cdot 
{\bf p}$ perturbation theory \cite{kpa} is 
$\sim ({\bf p}_{c,v} \cdot \hat {\bf K})^2$, where $\hat {\bf K}$ is the unit
vector in the direction of ${\bf K}$. This term, therefore, 
is responsible for the non-analytic behavior of the energy of dipole-allowed
excitons at the zone center, i.e., $K \rightarrow 0$, depending on the relative 
direction of ${\bf K}$ with respect to ${\bf p}_{c,v}$ \cite{Bassani}.
The exchange interaction in this case lifts the degeneracy between longitudinal
$({\bf K} \parallel {\bf p}_{c,v}$) and transverse $({\bf K} \perp 
{\bf p}_{c,v}$)  excitons, with the longitudinal lying higher than the 
transverse. The same phenomenon appears in optical phonons, i.e., the 
longitudinal modes have higher energy than
the transverse at $K=0$, because
in the case of longitudinal oscillations there is charge
accumulation (not present in the case of transverse oscillations),
which creates an internal electric field \cite{phonons}.

    By contrast, in Cu$_2$O, the second term of Eq. (\ref{222}) ($\sim K^2$) 
makes the 
total mass of the ortho excitons smaller than $m_e+m_h$; for the para excitons
it vanishes. To estimate this correction, we
use ${\bf k} \cdot {\bf p}$ perturbation theory to write
\begin{eqnarray}
       |u_{l, {\bf q} + {\bf K}/2} \rangle
      \approx |u_{l, {\bf q}} \rangle +
     \frac {\hbar} m \sum_{n \ne l} \frac {\langle u_{n, {\bf q}}|
    ({\bf K}/2) \cdot {\bf p} \,|
   u_{l,{\bf q} } \rangle } { \varepsilon_{l, {\bf q} } -
  \varepsilon_{n, {\bf q} } }
 |u_{n, {\bf q} } \rangle,
\label{62}
\end{eqnarray}
where the sum is over all the bands of parity opposite to that of band 
$l$. Thus, for the sums in the second term of Eq. (\ref{222}),
\begin{eqnarray}
    \sum_{{\bf q}} \phi_{{\bf q}} 
      \langle u_{c,{\bf q}+ {\bf K}/2}|
        u_{v, {\bf q} - {\bf K}/2}\rangle \approx
        \left( \frac \hbar {2m} \right)^2 \left( \sum_n  \frac
        {\langle u_{c, 0 }| {\bf K} \cdot {\bf p} \, |
      u_{n, 0 } \rangle \langle u_{n, 0 }| {\bf K} \cdot {\bf p} \, |
      u_{v, 0 } \rangle} {(\varepsilon_{c, 0 } -
    \varepsilon_{n, 0 }) (\varepsilon_{n, 0 } -
    \varepsilon_{v, 0 })}\right) \Omega_c^{1/2} \Phi_{1s}(0),
\label{63}
\end{eqnarray}
where the sum is over the two negative-parity bands of Cu$_2$O.
This correction modifies the free dispersion relation of the ortho excitons to
\begin{eqnarray}
     E_{{\bf K}} = \frac {\hbar^2 K^2} {2 M}
       \left[ 1+ \frac 1 3 \frac M {\mu} 
      \frac {\epsilon_{0}} {\epsilon_{\infty}}
     \left( \frac {\hbar} {m a_B} \right)^4
    \sum_n 
   \frac {(\hat {\bf K} \cdot {\bf p}_{c,n})^2 (\hat {\bf K} \cdot 
   {\bf p}_{n,v})^2}
  {(\varepsilon_{c,0} - \varepsilon_{n,0})^2
 (\varepsilon_{n,0} - \varepsilon_{v,0})^2} \right].
\label{233}
\end{eqnarray}
The above equation predicts that the ortho-exciton mass due to the correction
caused by the exchange interaction is not isotropic. 
Using the estimate $|{\bf p}_{i,j}|/\hbar \approx 0.13$ \AA$^{-1}$ and
the energy levels of Cu$_2$O, we find that
the correction [i.e., the second term in the parentheses in Eq. (\ref{233})]
of the exchange interaction to the (ortho) exciton mass
is negligibly small, $\approx 0.001$.

\subsection{Band structure of Cu$_2$O and optical properties}

      The $\Gamma_6^{+}$ conduction band in Cu$_2$O is formed by Cu 4s 
orbitals and the $\Gamma_7^{+}$ valence band by
Cu 3d orbitals \cite{bstr}. The fivefold degenerate (without spin) Cu 3d
orbitals split under the crystal field
into a higher threefold $\Gamma_{25}^{+}$ and a lower $\Gamma_{12}^{+}$
twofold degenerate band. Finally, $\Gamma_{25}^{+}$ splits further
because of the spin-orbit interaction into two bands, a higher 
$\Gamma_{7}^{+}$ non-degenerate band and a lower twofold degenerate 
$\Gamma_{8}^{+}$ band (Fig. 1).

    The total angular momentum functions for the yellow-exciton
triplet states are
\begin{eqnarray}
    &|J=1, J_z=1 \rangle& =  |\uparrow_e, \uparrow_H \rangle
\\
        &|J=1, J_z=0 \rangle& =
       \frac 1 {\sqrt 2}  (|\uparrow_e, \downarrow_H \rangle -
      |\downarrow_e, \uparrow_H \rangle ) 
\\
        &|J=1, J_z=-1 \rangle& = |\downarrow_e, \downarrow_H \rangle,
\label{238}
\end{eqnarray}
and for the singlet states,
\begin{eqnarray}
     |J=0, J_z=0 \rangle = \frac 1 {\sqrt 2}
     (|\uparrow_e, \downarrow_H \rangle +
   |\downarrow_e, \uparrow_H \rangle)  .
\label{239}
\end{eqnarray}
The indices $e,H$ refer to the electron and the hole, respectively: while
the electron states are pure spin states, the hole states are {\it total\/}
angular momentum states,
\begin{eqnarray}
    |\uparrow_H \rangle = - \frac 1 {\sqrt 3} \lbrack (X+iY)
\, |\downarrow_h \rangle
  + Z \, |\uparrow_h \rangle \rbrack
\label{240}
\end{eqnarray}
and
\begin{eqnarray}
     |\downarrow_H \rangle = - \frac 1 {\sqrt 3} \lbrack (X-iY)
\,      |\uparrow_h \rangle
        - Z \, |\downarrow_h \rangle \rbrack ,
\label{241}
\end{eqnarray}
where the states with lower case $h$
are pure spin-states. The spatial functions
$X,Y,Z$ transform as yz, xz and xy, respectively.

    The above angular momentum functions explain why the exchange
interaction is nonzero only for the ortho excitons, as well as why 
the direct recombination of the para excitons is highly forbidden.
If we assume that there is no spin-flip (a higher-order effect),
the above angular momentum functions imply that the exchange diagram
shown in Fig. 3 (virtual annihilation of the exciton) vanishes for the singlet.
For the triplet state the exchange interaction does not vanish and
raises the ortho-exciton with respect to the para-exciton energy.
The factor of $2/3$ we used for the calculation of the exchange energy
in Eq. (\ref{221}) comes from the above angular momentum functions.
The radiative recombination process of excitons is essentially described by 
the left (right) half of the virtual annihilation diagram,
with the only difference that a real instead of a virtual photon is 
emitted. The matrix element for the recombination process, therefore,
is proportional to $\sqrt {2/3}$ times the result from the spatial part
of the calculation for the ortho excitons, but it vanishes for the
para excitons. The same physics is responsible for the ortho excitons 
lying higher than the para excitons, and the direct recombination of the
para excitons being highly forbidden.

\section{Summary}

    Based on the effective mass approximation, we have used perturbation 
theory and the variational method to calculate the binding energy, the
Bohr radius as well as the total mass of the 1s-state of the yellow-exciton 
series. We have shown that the non-parabolicity of the bands gives 
consistent corrections for the total exciton mass and the exciton binding 
energy, as well. The coupling of the electrons and the holes to the LO-phonons
and the momentum dependence of 
the dielectric function also contribute to the binding energy.
The exchange interaction is responsible for the energy splitting between 
the triplet and the singlet-state excitons at the zone center, with the
triplet lying higher because of the band structure of Cu$_2$O.
Finally, the contribution of the exchange interaction to the exciton mass  
is negligible.

\acknowledgments
     This work was supported by NSF Grant No. PHY94-21309.
     Helpful comments from 
     K. O'Hara, L. O'Suilleabhain, D.W. Snoke, and J.P. Wolfe are gratefully
  acknowledged. G.M.K. would like to thank the Research Center of Crete, Greece
     for its hospitality.

\figure{FIG. 1.
     Schematic band structure of Cu$_2$O
showing the conduction $\Gamma_6^{+}$ band
and the $\Gamma_7^{+}$, $\Gamma_8^{+}$ valence bands, split by the spin-orbit
splitting, which form the yellow and green-exciton series, respectively.}
 
\figure{FIG. 2. The solid line shows
  the expectation value of the energy of the 1s-exciton in Cu$_2$O
  as function of the Bohr radius, Eq. (\ref{total}); the dotted line
  shows the same function with the central-cell corrections 
  not taken into account.}

\figure{FIG. 3.
     The virtual annihilation of an exciton, possible only for
     pure spin-singlet states.}

\end{document}